\documentclass[aps,pra,preprint,amsmath,amssymb]{revtex4-1}
\usepackage[T1]{fontenc}
\usepackage{graphicx}
\usepackage{dcolumn}
\usepackage{bm}
\usepackage{subfigure}

\begin{document}

\preprint{APS/123-QED}

\title{Strong-coupling expansion for the spin-1 Bose--Hubbard model}


\author{Takashi Kimura}
\email[]{tkimura@kanagawa-u.ac.jp}
\affiliation{Department of Mathematics and Physics, Kanagawa University, 2946 Tsuchiya, Hiratsuka, Kanagawa 259-1293, Japan}


\date{\today}

\begin{abstract}
In this study, we perform a strong-coupling expansion up to third order of
the hopping parameter $t$ for the spin-1 Bose--Hubbard model 
with antiferromagnetic interaction. 
As expected from previous studies,  
the Mott insulator phase 
is considerably more stable 
against the superfluid phase 
when filling with an even number of bosons 
than when filling with an odd number of bosons. 
The phase-boundary curves are consistent 
with the perturbative mean-field theory in the 
limit of infinite dimensions. 
The critical 
value of the hopping parameter $t_{\rm C}$
at the peak of the Mott lobe depends on the antiferromagnetic interaction. 
This result indicates  the reliability of the 
strong-coupling expansion when $U_2$ possesses large (intermediate) values 
for Mott lobe with an even (odd) number of bosons. 
Moreover, in order to improve our results, we apply a few extrapolation 
methods up to infinite order in $t$. 
The fitting results of the phase-boundary curves
agree better with those of the perturbative 
mean-field approximation. 
In addition, the linear fit error of $t_{\rm C}$ is very small
for the strong antiferromagnetic interaction.  
\end{abstract}


\maketitle

\section{INTRODUCTION}
Since the realization of the Bose--Einstein condensation,  
ultracold bosons have been extensively studied. In 
trapped-atom systems, 
the temperature can reach approximately zero, 
which is very difficult to realize in 
conventional experimental systems. 
In addition to conventional spinless bosons, 
spinor bosons have also been examined as a new bosonic system 
with multiple internal degrees of 
freedom \cite{Stamper-Kurn1,Stamper-Kurn2}. 
The development of optical lattice systems 
has further promoted the study of ultracold bosons. In particular, 
 the transition from superfluid (SF) to Mott insulator (MI) 
has been obtained in an optical lattice system 
\cite{Greiner,Bakr1,Bakr2,Endres,Weitenberg}.

In theory, an optical lattice system with low boson filling
can generally be described by 
the Bose--Hubbard (BH) model \cite{Fisher,Jaksch}. 
In addition, both the MI phases and SF--MI transitions of spin-1 bosons 
have been intensively studied 
\cite{Demler,Yip,Imambekov,Snoek,Zhou,Uesugi,Tsuchiya,Kimura1,Krutitsky1,Krutitsky2,Kimura2,Yamashita,Rizzi,Bergkvist,Batrouni,Apaja,Toga,Lacki,Wagner}. 
The ferromagnetically interacting system is essentially similar to 
spinless bosons, whereas the antiferromagnetically interacting system exhibits
rich physical properties. 
Several spin phases such as 
the singlet, nematic, and dimerized phases in the insulating phase 
have also been 
analytically \cite{Demler,Yip,Imambekov,Snoek,Zhou}
and numerically \cite{Rizzi,Bergkvist,Batrouni,Apaja}
examined. 
To study the SF--MI transition, Tsuchiya {\it et al}.
\cite{Tsuchiya} used perturbative mean-field approximation 
(PMFA) \cite{Oosten}, 
which expands the free energy in the SF order parameter, 
to quantitatively show that 
the MI phase for filling with an even number of bosons 
(hereafter ``even boson filling'') is considerably more 
stable against the SF phase than that 
for filling with an odd number of bosons 
(hereafter ``odd boson filling''). 
This conjecture has been confirmed by the density matrix renormalization 
group (DMRG) \cite{Rizzi} and 
quantum Monte Carlo simulation (QMC) \cite{Apaja,Batrouni} 
in one dimension (1D).  However, 
 mean-field (MF) studies beyond the perturbation 
theory \cite{Kimura1,Krutitsky1,Krutitsky2} 
have shown a possible first-order 
SF--MI transition of the BH model 
for a weak antiferromagnetic interaction, such as 
$U_2/U_0\simeq 0.04$, which corresponds to ${}^{23}$Na. 
This first-order transition in 1D has also been revealed 
by a QMC study \cite{Batrouni}. 
However, if the antiferromagnetic interaction 
is adequately strong, the first-order transition may be neglected 
because it occurs when kinetic energy is considerably greater than 
antiferromagnetic-interaction energy near the SF-MI phase boundary 
\cite{comment0}. 
For a second-order transition, 
strong-coupling expansion of kinetic energy \cite{Freericks}, which is based on 
the Rayleigh--Schr{\"o}dinger perturbation 
theory \cite{Cohen}, is an excellent method for obtaining the phase boundary. 
The strong-coupling expansion has been applied to the analysis of the 
spinless \cite{Freericks,Kuhner,Buonsante,Sengupta,Freericks2,Varma}, 
extended \cite{Iskin}, hardcore \cite{Hen}, 
and two-species models \cite{Iskin2},   
and the results agree very well 
 with QMC results \cite{Freericks2,Hen}. 
To date, however, only MF calculations have been performed to 
analytically study the SF--MI transition of the spin-1 BH model.

In this study, we perform a strong-coupling expansion 
of the spin-1 BH model up to the third order of the hopping parameter $t$. 
The rest of this paper is organized as follows:  
Section II introduces the spin-1 BH model and strong-coupling expansion.
Section III provides the results: 
the phase diagrams for two dimensions (2D) or three dimensions (3D); 
the critical values of $t$ at the peak of the Mott lobes 
and their dependence on antiferromagnetic interaction,  
which reflect the validity of the expansion; 
results obtained by several extrapolation techniques that go up to 
the infinite order of $t$; and the 1D phase diagram. 
A summary of the results and discussions are given in Sec. V. 

\section{SPIN-1 BOSE--HUBBARD MODEL}
The spin-1 BH model 
is given by $H=H_0+H_1$,
\begin{eqnarray}
H_0&=&-t\sum_{\langle i,j\rangle,\alpha}(a_{i \alpha}^\dagger a_{j \alpha}^{}
 + a_{j \alpha}^\dagger a_{i \alpha}^{}), \nonumber\\
H_1&=&-\mu\sum_{i,\alpha} a_{i \alpha}^\dagger a_{i \alpha}^{} 
+\frac{1}{2}U_0\sum_{i,\alpha,\beta} a_{i \alpha}^\dagger a_{i \beta}^\dagger
a_{i \beta}^{}a_{i \alpha}^{}\nonumber\\
&&+\frac{1}{2} U_2 \sum_{i,\alpha,\beta,\gamma,\delta} 
a_{i \alpha}^\dagger a_{i \gamma}^\dagger
{\bf F}_{\alpha \beta} \cdot {\bf F}_{\gamma \delta} a_{i \delta}^{}
 a_{i \beta}^{}.\label{H}\nonumber\\
&=&\sum_i\Big[-\mu\hat{n}_i+\frac{1}{2}U_0\hat{n}_i(\hat{n}_i-1)
+\frac{1}{2}U_2({\hat{\mathbf S}_i}^2-2\hat{n}_i)\Big].
\end{eqnarray}
Here, $\mu$ and $t(>0)$ are the chemical potential 
and the hopping matrix element, respectively.  
The quantity $U_0$ ($U_2$) is the spin-independent (dependent) 
interaction between bosons. We assume that $U_0$ and $U_2$ are 
positive, which correspond to repulsive and antiferromagnetic interaction, 
respectively. 
The operator $a_{i \alpha}$ ($a_{i \alpha}^\dagger$) 
annihilates (creates) 
 a boson at site $i$ with spin-magnetic quantum number 
$\alpha=1,0,-1$. The number operator at site $i$ is given by 
$n_i\equiv\sum_\alpha n_{i\alpha}$ ($n_{i\alpha}\equiv 
a_{i \alpha}^\dagger a_{i \alpha}$). 
The spin operator at site $i$ is 
$ {\hat{\bf S}_i}\equiv \sum_{\alpha,\beta}a_{i \alpha}^\dagger 
{\bf F}_{\alpha \beta} a_{i \beta}$ and 
${\bf F}_{\alpha \beta}$ represents the spin-1 matrices. 
In this study, we assume a tight-binding model 
with only nearest-neighbor hopping. 
The summation over all sets of adjacent sites 
is expressed by $\langle i,j \rangle$. 
For simplicity, we assume a hypercubic lattice. 

Under the limit $t\rightarrow 0$, 
the MI phase exists for arbitrary $\mu$;
the MI phase also has an even number of bosons
$n_0$ per site for $U_0(n_0-1)-2U_2<\mu<U_0n_0$ 
or an odd number of bosons $n_0$ for $U_0(n_0-1)<\mu<U_0n_0-2U_2$. 
To ensure that 
the phase diagram has MI phases with an odd number of bosons per site, 
we assume $U_0> 2U_2>0$. 
The SF--MI phase boundary can be determined by calculating  
the  energy of the MI phase and that of the defect state,  
which has exactly one extra particle or hole. 
Specifically, if $E_{\rm MI}(n_0,\mu,t)> (<) \min\Big(E^{\rm part}(n_0,\mu,t), 
E^{\rm hole}(n_0,\mu,t)\Big)$, then 
the phase is SF (MI),  
where $E_{\rm MI}(n_0,\mu,t)$ is the energy of the MI state 
and $E^{\rm part}(n_0,\mu,t)$ [$E^{\rm hole}(n_0,\mu,t)$]
is the energy of the defect state with one extra particle (hole).
The SF--MI phase boundary is determined by
\begin{eqnarray}
  E_{\rm MI}(n_0,\mu,t)&=&E^{\rm part}(n_0,\mu,t)
\end{eqnarray}
or
\begin{eqnarray}
  E_{\rm MI}(n_0,\mu,t)&=&E^{\rm hole}(n_0,\mu,t). 
\end{eqnarray}

\section{STRONG-COUPLING EXPANSION}
Following Ref. \cite{Freericks}, we employ 
the Rayleigh-Schr{\"o}dinger perturbation theory 
for calculations up to the third order of the hopping parameter $t$. 
We start from the unperturbed MI states, 
define the defect state by doping an extra particle or hole into the MI states, 
and compare the energy of the MI state with that of the defect state. 

\subsection{Mott-insulator states at the zeroth order of the hopping parameter}
The unperturbed MI wave function with an even number of bosons per site is 
\begin{equation}
\Psi_{\rm even}=\prod_k |n_0,0,0\rangle_k, \label{Me}
\end{equation}
where $|n_0,0,0\rangle_k$
implies the boson number $N=n_0$, the spin magnitude 
$S=0$, and the spin magnetic quantum number $S_z=0$ at site $k$. 
For simplicity, we neglect the nematic MI state that 
includes $S=2$ states. However, from analytical 
calculations \cite{Imambekov,Snoek}, 
the nematic MI phase for even boson filling may exist 
for a weak $U_2$. 

For the unperturbed MI state with an odd number of bosons per site, we assume 
a nematic MI state 
\begin{equation} 
\Psi_{\rm odd}=\prod_k |n_0,1,0\rangle_k.   \label{Mo}
\end{equation}
Although $\Psi_{\rm ferro}=\prod_k |n_0,1,\pm 1\rangle_k$
is degenerate with $\Psi_{\rm odd}$ at $t=0$, 
we can easily find that the degeneracy is lifted 
for finite $t$, and $\Psi_{\rm odd}$ has lower energy,  
at least up to the third-order perturbation of $t$, as expected. 
This is natural because we assume antiferromagnetic interaction. 
The dimerized state is also degenerate with 
$\Psi_{\rm odd}$ at $t=0$ 
and is considered to be the ground state for finite $t$ in 1D 
\cite{Yip,Imambekov,Zhou,Rizzi,Bergkvist,Apaja,Batrouni}. 
Therefore, the validity of the results 
based on $\Psi_{\rm odd}$ for odd boson filling 
is basically limited to 2D or 3D systems, although the 
existence of the dimerized phase 
cannot be denied even there. 
We note that $\Psi_{\rm even}$ ($\Psi_{\rm odd}$) is also adopted as 
the ground state in PMFA \cite{Tsuchiya}, 
which we compare with the results in the following section. 

\subsection{Defect states}
We define the defect states by doping an extra particle or hole into 
$\Psi_{\rm even}$ and $\Psi_{\rm odd}$ as follows: 
\begin{eqnarray}
\Psi^{\rm part}_{\rm even}&=&
\frac{1}{\sqrt{N}} \sum_{i}\Big[ f_i
|n_0+1,1,0\rangle_i\otimes\prod_{k\ne i} |n_0,0,0\rangle_k\Big],\label{pe}\\
\Psi^{\rm hole}_{\rm even}&=&\frac{1}{\sqrt{N}} \sum_{i}\Big[f_i
|n_0-1,1,0\rangle_i
\otimes\prod_{k\ne i} |n_0,0,0\rangle_k\Big],\label{he}\\
\Psi^{\rm part}_{\rm odd}&=&\frac{1}{\sqrt{N}} \sum_{i}\Big[ f_i
|n_0+1,0,0\rangle_i
\otimes\prod_{k\ne i} |n_0,1,0\rangle_k\Big],\label{po}\\
\Psi^{\rm hole}_{\rm odd}&=&\frac{1}{\sqrt{N}} \sum_{i}\Big[ f_i
|n_0-1,0,0\rangle_i
\otimes\prod_{k\ne i} |n_0,1,0\rangle_k\Big]. \label{ho}
\end{eqnarray} 
Here $N$ is the number of lattice sites, and $f_i$ is the 
eigenvector of the hopping matrix $t_{ij}$ 
with the highest eigenvalue \cite{Freericks}. 
In this study, because we assume hypercubic lattices with 
only nearest-neighbor hopping, $f_i=1$ and 
the eigenvalue $\lambda=zt$,  
where $z=2d$ is the number of nearest-neighbor sites in the $d$-dimensional 
hypercubic lattice.  Although, 
$\Psi^{\rm part(hole)}_{\rm odd}$ has no other degenerate 
candidates, $\Psi^{\rm part(hole)}_{\rm even}$ is 
degenerate with 
$\Theta^{\rm part(hole)}_{\pm}=\frac{1}{\sqrt{N}} \sum_{i}\big[ |n_0+(-) 1,1,\pm 1 \rangle_i\otimes\prod_{k\ne i} |n_0,1,0\rangle_k\big]$. 
We find that $\Theta^{\rm part(hole)}_{\pm}$ has the exact same energy 
as $\Psi^{\rm part(hole)}_{\rm even}$ up to the third order of $t$ 
and that we can choose $\Psi^{\rm part(hole)}_{\rm even}$ as the defect state. 
We note that $\Psi^{\rm part(hole)}_{\rm even}$ is nonmagnetic 
like the SF phase is nonmagnetic. 

\subsection{Ground-state energies and phase diagrams}
By using $\Psi_{\rm even(odd)}$ and $\Psi^{\rm part(hole)}_{\rm even(odd)}$ 
from Sec. III B,  
the energies of the MI state with an even or odd number of bosons 
per site
and those of the defect states 
are obtained up to the third order of $t$,   
as shown by Eqs. (\ref{evenenergy})--(\ref{oddhole})
of Appendix A. 

By equating the right-hand side of Eqs. (\ref{evenpart})--(\ref{oddhole})
to zero (where the MI and the defect states are degenerate), 
we obtain the SF--MI phase-boundary $t$--$\mu$ curves  
$\mu^{\rm part}_{\rm even}(t)$, $\mu^{\rm hole}_{\rm even}(t)$, 
$\mu^{\rm part}_{\rm odd}(t)$, and $\mu^{\rm hole}_{\rm odd}(t)$, 
which show the upper branch (corresponding to particle doping) or lower 
branch (corresponding to hole doping) of the phase-boundary curve 
around the Mott phase with even boson filling  
and odd boson filling, respectively. 

\begin{figure}
  \includegraphics[height=2.5in]{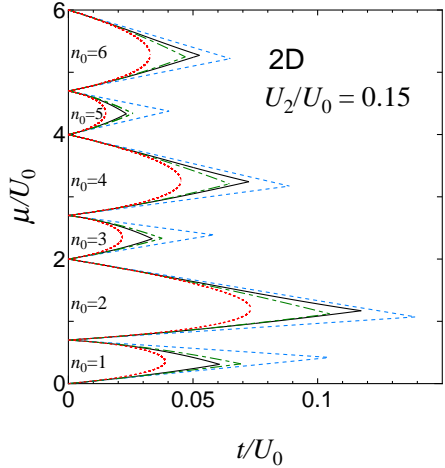}
  \caption{(Color online) 
Phase diagram obtained by the strong-coupling expansion
[Eqs. (\ref{evenpart})--(\ref{oddhole})] for $U_2/U_0=0.15$ in 2D. 
The solid curves show the results up to the third order 
of $t$. Results up to the first order (second order) of $t$ 
are also shown by the blue dashed (green dot-dashed) curve. 
The red dotted curve shows the results obtained by PMFA. 
}
\label{phasediagram2Dus015}
\end{figure}

\begin{figure}
  \includegraphics[height=2.5in]{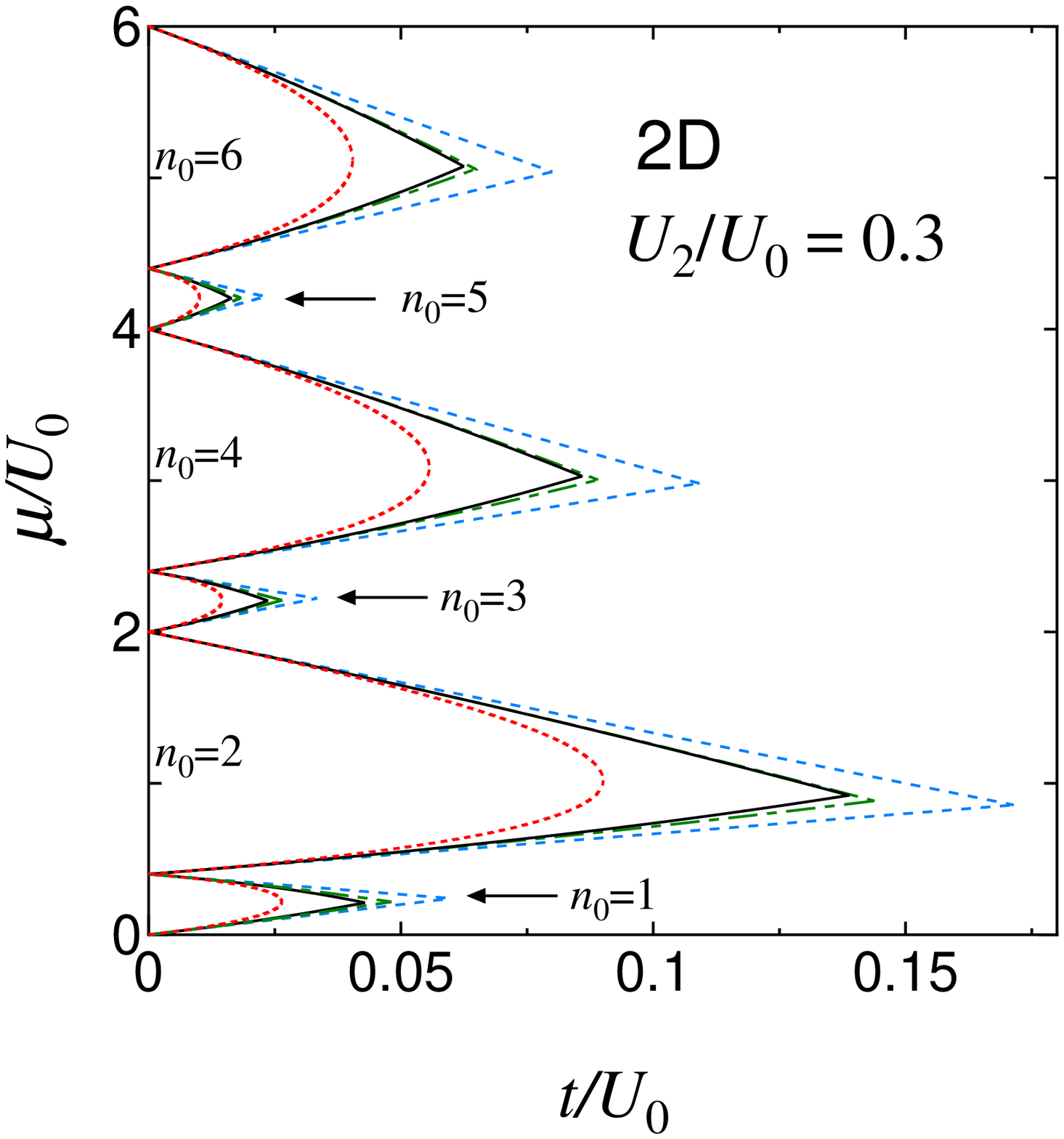}
  \caption{(Color online) 
Same plot as  in Fig. \ref{phasediagram2Dus015} but for $U_2/U_0=0.3$ in 2D. 
}
\label{phasediagram2Dus03}
\end{figure}

\begin{figure}
  \includegraphics[height=2.5in]{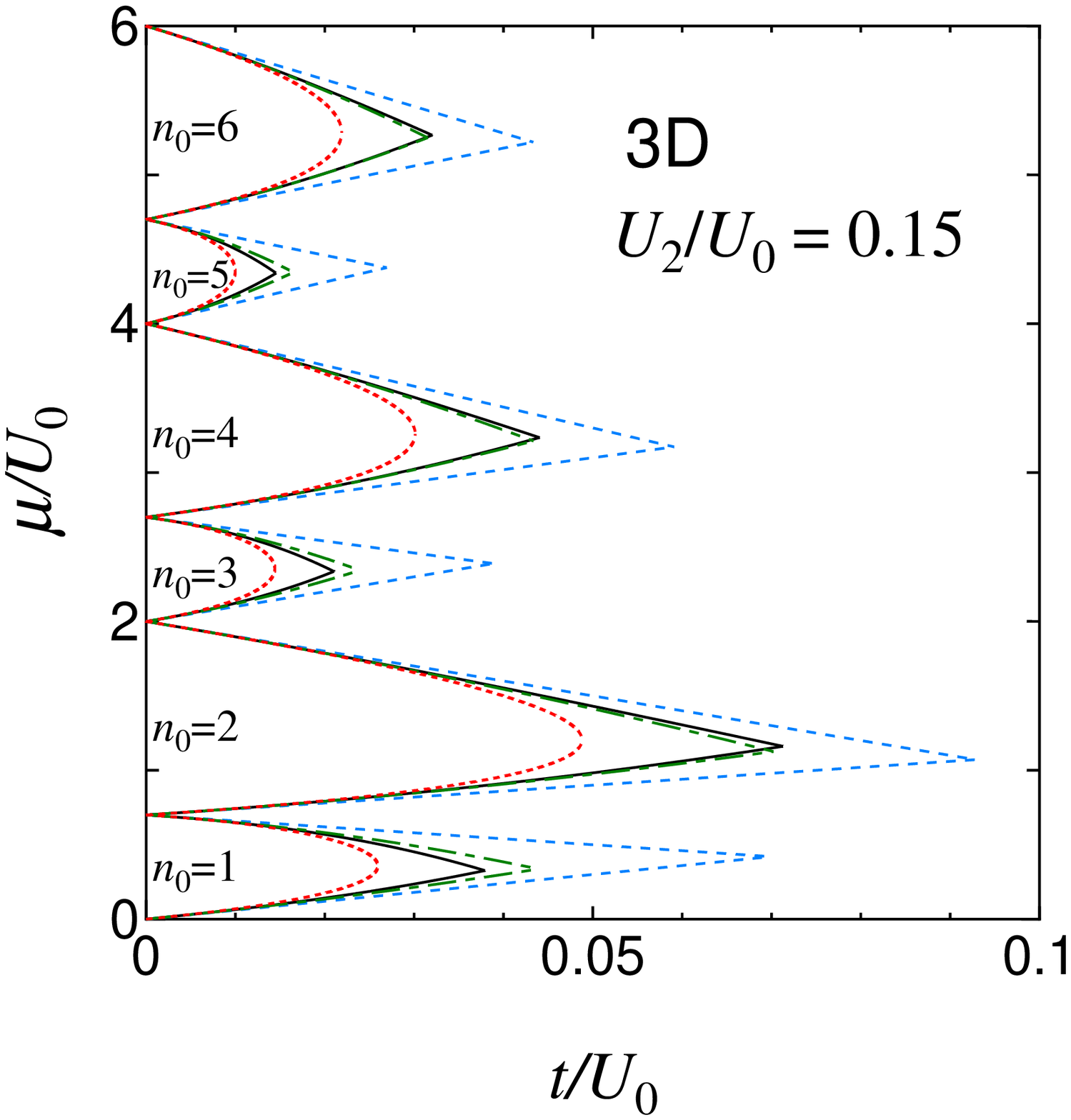}
  \caption{(Color online) 
Same plot as in Fig. \ref{phasediagram2Dus015} but for $U_2/U_0=0.15$ in 3D. 
}
\label{phasediagram3Dus015}
\end{figure}

\begin{figure}
  \includegraphics[height=2.5in]{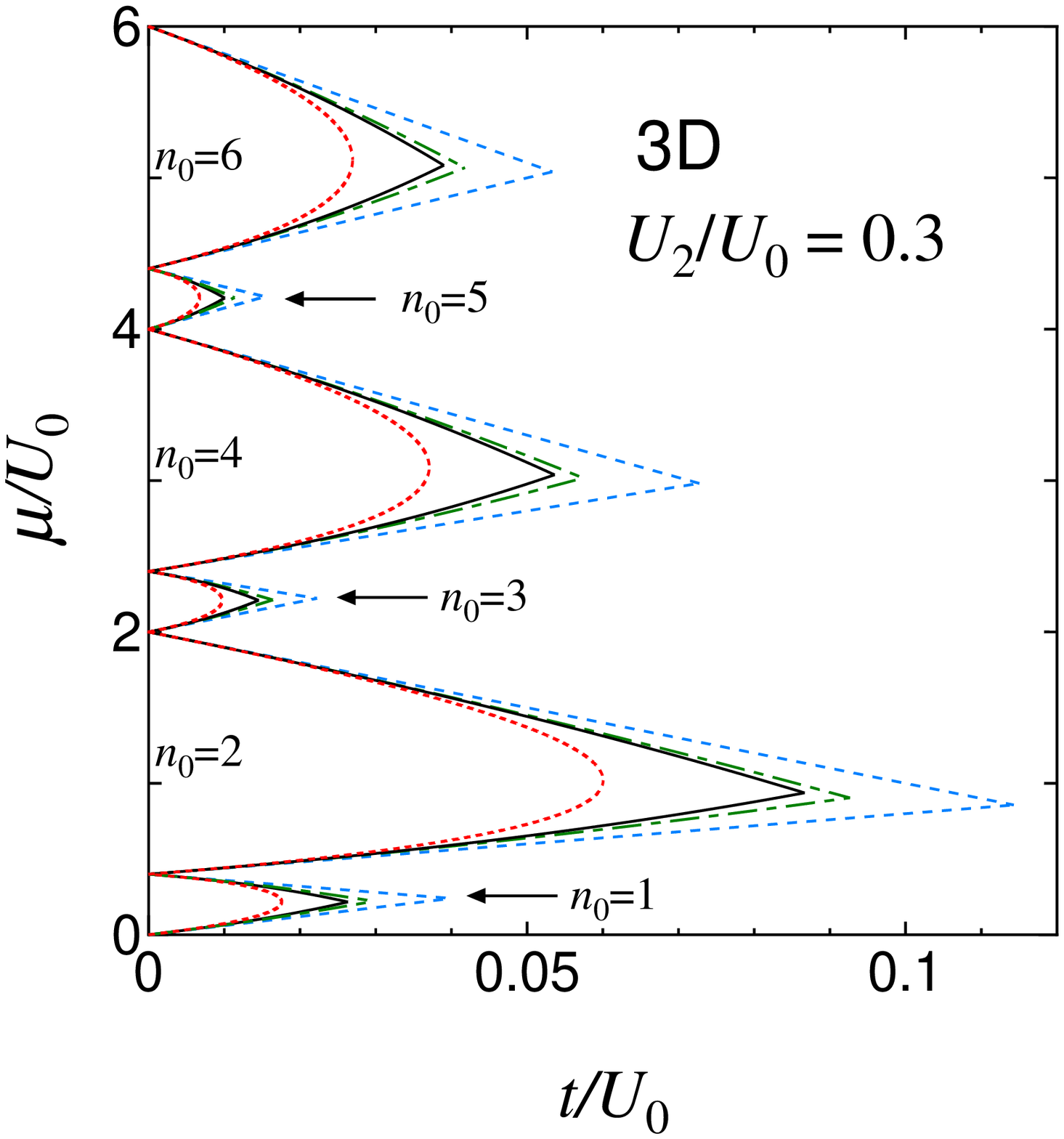}
  \caption{(Color online) 
Same plot as in Fig. \ref{phasediagram2Dus015} but for $U_2/U_0=0.3$ in 3D. 
}
\label{phasediagram3Dus03}
\end{figure}

Figures \ref{phasediagram2Dus015}--\ref{phasediagram3Dus03} 
show the phase diagram obtained from the calculation. 
The MI phase for even boson filling is considerably more stable 
against the SF phase than that for odd boson filling, as expected 
from MF and QMC studies. The area of the MI for even (odd) boson filling
increases more (decreases more) for $U_2/U_0=0.3$ than that for $U_2/U_0=0.15$. 
The critical value of $t$ on the phase-boundary curve 
is greater in 2D than that in 3D because the number of nearest-neighbor 
sites $z$ and the kinetic energy are smaller for a given value of $t$. 
These curves show the convergence of the strong-coupling expansion 
from the first to the third order of $t$, 
and we find that convergence is excellent 
except in Fig. \ref{phasediagram2Dus015}, where $U_2=0.15U_0$ is weak in 2D. 

In addition, we plot the results of 
PMFA; these results are exact in the limit of infinite dimensions,  
provided the SF--MI transition is of the second order. 
The area of the Mott lobes obtained by the strong-coupling expansion 
is greater than that obtained by PMFA. This difference 
may reflect the quantum fluctuations in the MI phases, 
which are incorporated (neglected) in the strong-coupling expansion (PMFA). 


\subsection{Consistency with PMFA}
PMFA involves MF decoupling theory using a 
perturbative expansion of the SF order parameter. 
If the SF--MI transition is of the second order, 
the phase-boundary curve obtained by PMFA is exact in infinite dimensions.  
Thus, if we expand the equation for this phase-boundary curve, 
obtained by PMFA 
up to the third order of $zt$, it must agree with the 
proposed strong-coupling expansion in infinite dimensions. 
The results of the expansion of 
this phase-boundary curve obtained by PMFA 
[Eqs. (30) and (46) of Ref. \cite{Tsuchiya}] 
are given by Eqs. (\ref{muevenpart})--(\ref{muoddhole}) of 
Appendix B. The results are consistent with the proposed 
strong-coupling expansion. 
Specifically, we find that the solutions of the equations 
$E_{\rm MI,even(odd)}(n_0)=E^{\rm part(hole)}(n_0)_{\rm def,even(odd)}$ 
are the same as Eqs. (\ref{muevenpart})--(\ref{muoddhole}) 
under the limit $z\rightarrow\infty$ and $t\rightarrow 0$ with 
constant $zt$.  

\subsection{Critical value of $t$ at the peak of the Mott lobe}
In this section, we examine the critical value $t_{\rm C}$ 
of the hopping parameter 
$t$ at the peak of the Mott lobe, where 
the upper branch of the $t$--$\mu$ 
curve for the phase-boundary 
converges with its lower branch. In addition, 
the dependence of $t_{\rm C}$ on $U_2$ 
indicates the range over which 
the proposed expansion up to  the third order of $t$ 
may be applied. 

\begin{figure}
  \includegraphics[height=2.5in]{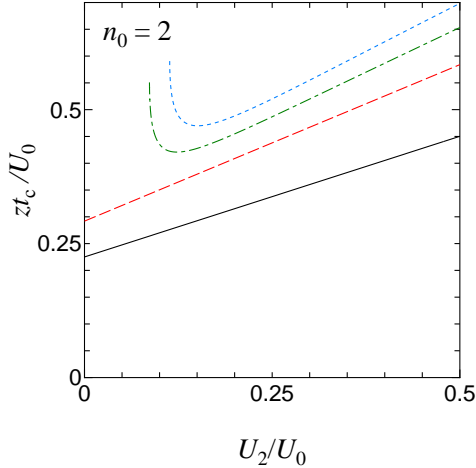}
  \caption{(Color online) 
Dependence of $t_{\rm C}$ on $U_2$ 
for Mott lobe with even boson filling ($n_0=2$).  
The blue short-dashed, green dot-dashed, 
and red long-dashed curves show the results 
obtained by the strong-coupling expansion up to the 
third order of $t$ in 2D, 3D, and infinite dimensions, respectively. 
The solid curves show the results obtained by PMFA. 
}
\label{ztcu2depn2}
\end{figure}

Figure \ref{ztcu2depn2} shows 
the dependence of $t_{\rm C}$ on $U_2$ 
for a Mott lobe with an even number of bosons ($n_0=2$). 
The curves in infinite dimensions obtained by PMFA and 
those obtained by the strong-coupling expansion up to the third order of $t$
(see previous Sec. III D) 
 are smoothly increasing functions of $U_2$.  
When $U_2/U_0$ is large,  
the results obtained 
by the strong-coupling expansion up to the third order of $t$
in 2D and  3D show a similar dependence of $t_{\rm C}$ on $U_2$. 
However, the results show a strange behavior for $U_2/U_0\sim 0.1$, 
and the curves stop at small $U_2/U_0$ because we cannot 
find $t_{\rm C}$ as the upper and lower branches of the $t$--$\mu$ 
curve no longer converge. Such a situation also occurs for greater $n$. 
Because $U_2$ stabilizes the MI phase with even boson filling 
against the SF phase, the results of PMFA and of the 
strong-coupling expansion in infinite dimensions agree with physical intuition. 
The small $U_2$ regime is hazardous 
for the strong-coupling expansion in finite dimensions because 
a few denominators of the expansion 
contain $3U_2$, which corresponds to the spin-excitation energy 
[$E(S=2)-E(S=0)$] of an intermediate state that appears 
in perturbative calculation. 
This problem can be solved only by expanding  
$\mu$ to infinite orders of $t$. 
However, the terms involving $3U_2$ disappear in the denominators 
of the strong-coupling expansion in infinite dimensions, so we 
obtain $t_{\rm C}$ even for small $U_2/U_0$. 
On the other hand, the possibility of a first-order transition  
should not be ignored for a small $U_2/U_0$ \cite{Kimura1}. 

\begin{figure}
  \includegraphics[height=2.5in]{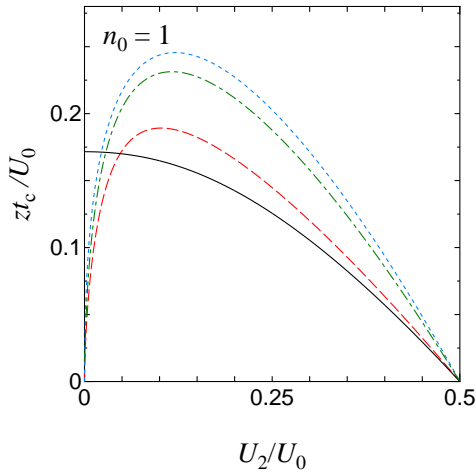}
  \caption{(Color online) 
Same plot as in Fig. \ref{ztcu2depn2} but 
for Mott lobe with odd boson filling ($n_0=1$).  
}
\label{ztcu2depn1}
\end{figure}

Figure \ref{ztcu2depn1} 
shows the same plot  as Fig. \ref{ztcu2depn2} but 
for a Mott lobe with an odd number of bosons ($n_0=1$). The parameter 
$t_{\rm C}$ obtained by PMFA 
is a smoothly decreasing function of $U_2$
and goes to zero at $U_2/U_0=0.5$, where the 
Mott lobes for odd boson filling disappear. Moreover, 
the other curves also show that $t_{\rm C}$ is a 
decreasing function of $U_2$ when $U_2/U_0$ is large.  
However, $t_{\rm C}$ is a  
rapidly increasing function of $U_2$ for $U_2/U_0 < 0.1$ and 
$t_{\rm C}=0$ at $U_2/U_0=0$. 
This is also because, in the strong-coupling expansion 
up to the third order of $t$,  
a few denominators contain $3U_2$
(here, this is true not only in 2D or 3D but also in infinite dimensions). 
Therefore, $t_{\rm C}\rightarrow 0$ when $U_2/U_0\rightarrow 0$, so that 
the two branches of the $t$--$\mu$ curve converge. 
For greater $n_0$, $t_{\rm C}$ cannot be obtained for a small $U_2/U_0$ 
for the same reason as that given in the previous paragraph 
for the case of even $n_0$ (figure not shown). 
A similar problem can also occur for $U_2/U_0\approx 0.5$ in 1D 
because a few terms in the expansion 
include $U_0-2U_2$ in their denominator. 
This problem can also be solved only by expanding  
$\mu$ to infinite orders of $t$. 
 
In summary, because the parameter $t_{\rm C}$ may be 
an increasing (decreasing) function of $U_2$ 
for a Mott lobe with even (odd) boson filling, 
the proposed strong-coupling expansion is reliable for 
large (intermediate) values of $U_2$.  

\subsection{Extrapolation methods}
The expansion up to the third order of $t$ 
has a few problems. For example, 
the phase-boundary curve, including 
the value of $t_{\rm C}$, does not completely converge.  
From a qualitative point of view, 
the expansion does not provide an appropriate scaling 
form of the phase-boundary curve near $t_{\rm C}$. 
However, in order to improve the phase diagram, 
we attempt two extrapolation methods in working 
toward an infinite-order theory in this section. 
\subsubsection{Linear fit of $t_{\rm C}$}
Critical-point extrapolation, 
which was proposed in Ref. \cite{Freericks},   
is a simple method that involves a 
least-squares fit to obtain a straight line 
that best fits the data. 

\begin{figure}
  \includegraphics[height=2.5in]{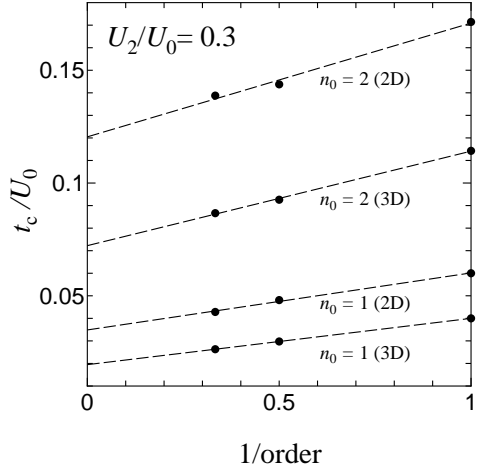}
  \caption{Solid circles show $t_{\rm C}$ obtained by the proposed
strong-coupling expansion up to the first, second, and third order of $t$
for $U_2/U_0=0.3$ in 2D and 3D. 
The data are labeled 
for $n_0=1$ Mott lobes and $n_0=2$ Mott lobes. 
The horizontal axis is the inverse of the order of expansion 
($1/{\rm order}=1/3,1/2,1$ for third, second, and first orders, respectively). 
The dashed lines are least-square linear fits of the solid circles. 
}
\label{tcfit} 
\end{figure}

Figure \ref{tcfit}  shows the 
critical point $t_{\rm c}$ at the peak of the Mott lobe 
for each order $m$ of the strong-coupling expansion.
The data for $t_{\rm C}$ lie approximately on a straight line. 
The data can be extrapolated to the infinite order ($1/m\rightarrow 0$)
by least-squares fitting to obtain the straight line 
that best fits the data. 
Specifically, extrapolating the straight line to 
where it intersects the vertical axis gives 
the infinite-order  $t_{\rm C}$. 

Table \ref{table1} gives the infinite-order fitting data. 
The fitting error \cite{comment1} is very small when $U_2/U_0$ is 
large, where strong-coupling expansion can be more reliable 
compared to the small-$U_2/U_0$ regime (which is consistent with the  
dependence of $t_{\rm C}$ on $U_2$ discussed in Sec. III E). 

\subsubsection{Extrapolation of phase-boundary curves}
\begin{figure}
  \includegraphics[height=2.5in]{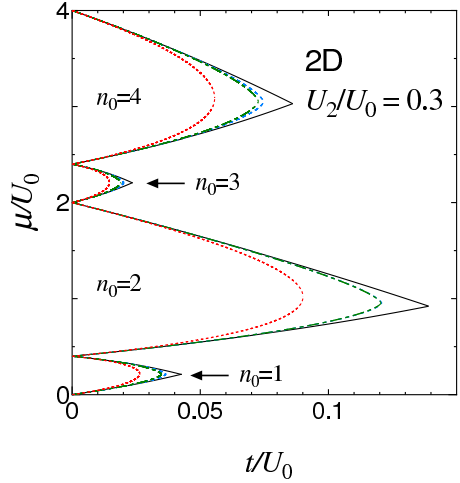}
  \caption{(Color on line)
Phase diagram obtained by strong-coupling expansion
up to the third order of $t$ (solid curve) for $U_2/U_0=0.3$ in 2D 
and its extrapolation to the infinite order of $t$. 
The blue dashed (green dot-dashed) curve shows the chemical-potential fitting 
without (with) the least-squares fit for $t_{\rm C}$. 
The blue dashed and green dot-dashed curves are similar to each other. 
In particular, 
the two curves for the $n_0=2$ Mott lobe are indistinguishable. 
The red dotted curve shows the PMFA results. The smallest area 
Mott lobe is obtained by PMFA. 
}
\label{fit2D} 
\end{figure}

\begin{figure}
  \includegraphics[height=2.5in]{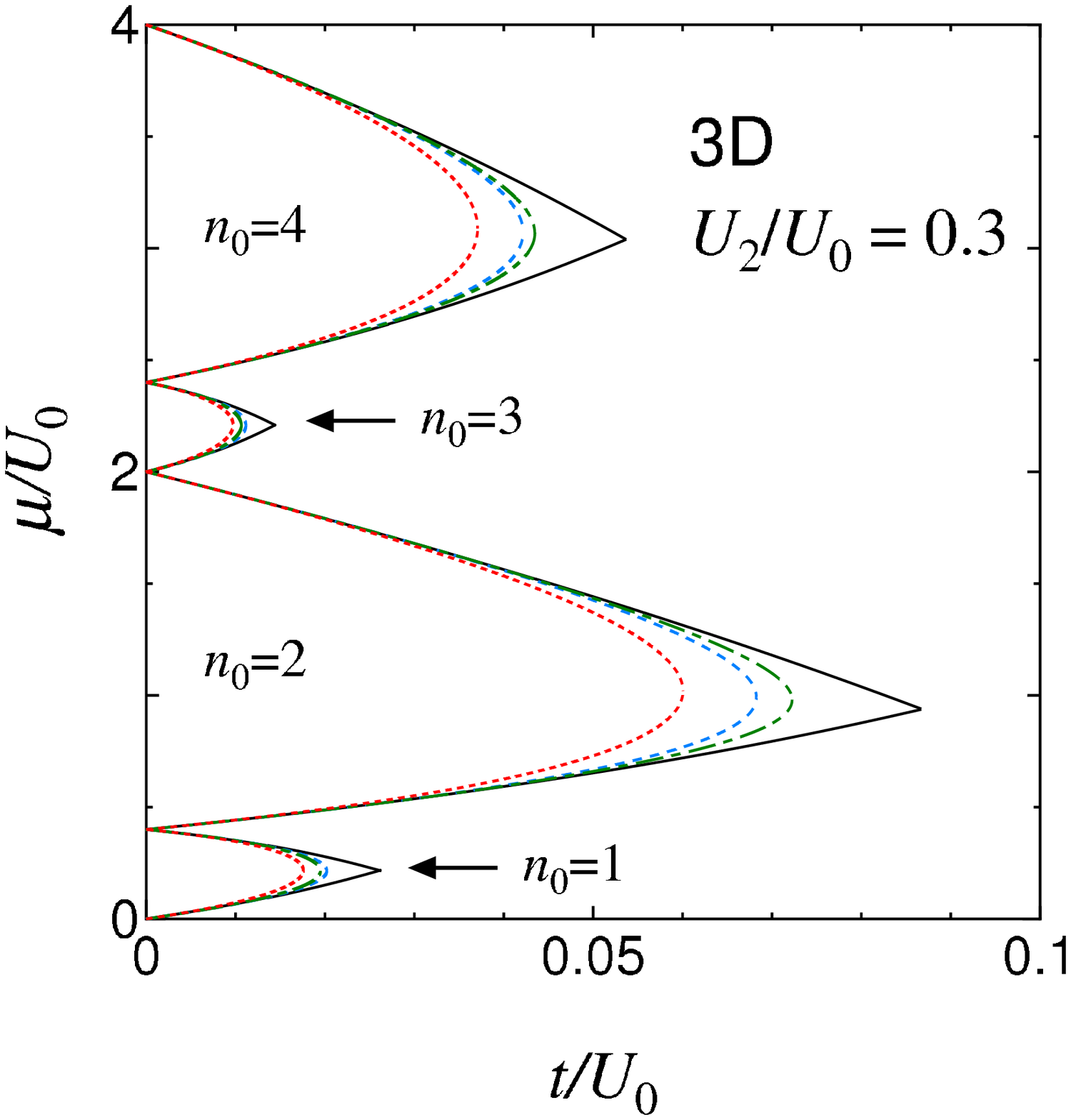}
  \caption{Same plot as that in Fig. \ref{fit2D} for 3D. 
}
\label{fit3D} 
\end{figure}

The proposed phase-boundary curve obtained by 
expansion up to the third order of 
$t$ has a cusp at the peak of the Mott lobe. 
However, it can be assumed that the chemical potential 
has the following power-law scaling near $t_{\rm C}$, 
similar to that of the spinless BH model in 2D or 3D:
\begin{eqnarray}
\mu=A(t)\pm B(t)(t_{\rm C}-t)^{z\nu}. 
\end{eqnarray}
The following fitting method is called chemical-potential fitting 
\cite{Freericks,Iskin}.
Here $A(t)\approx a+bt+ct^2+dt^3$ and
$B(t)\approx \alpha+\beta t+\gamma t^2+\delta t^3$ 
are the regular functions of $t$. 
The parameter $z\nu$ is the critical exponent in the model. 
By using expansion up to the third order
of $t$, we immediately determine $a$, $b$, $c$, and $d$ 
by setting $A(t)=[\mu^{\rm part}(t)+\mu^{\rm hole}(t)]/2$. 
In addition, by assuming that the scaling is 
the same as that of the spinless BH model \cite{Fisher,Freericks}, 
we set $z\nu\simeq 2/3$ for $d=2$ and  
$z\nu=1/2$ for $d>2$ in the $d$-dimensional spin-1 BH model 
\cite{comment2}. 
By setting $\delta=0$, we obtain $\alpha$, $\beta$, $\gamma$, 
and $t_{\rm C}$ by comparing the Taylor expansion of $t$ in 
$B(t)(t_{\rm C}-t)^{z\nu}$ 
with $[\mu^{\rm part}(t)-\mu^{\rm hole}(t)]/2$. 
The results for $t_{\rm C}$ are given in Table \ref{table1},  
and the phase-boundary curves obtained by the above fitting 
are shown in Fig. \ref{fit2D} for 2D and in 
Fig. \ref{fit3D} for 3D. 

By combining  these two fitting methods, we obtain the phase-boundary curve. 
Specifically, we use the value of $t_{\rm C}$ 
obtained by the least-squares fit to compare $B(t)(t_{\rm C}-t)^{z\nu}$ 
with $[\mu^{\rm part}(t)-\mu^{\rm hole}(t)]/2$. 
Here we include the $\delta t^3$ term in $B(t)$. The obtained 
phase-boundary curves are also plotted in Figs.  
\ref{fit2D} and \ref{fit3D}. 
The phase-boundary curves obtained by these two fitting methods 
are very similar, especially in 2D. 
The phase-boundary curves obtained by these two fitting methods are more 
similar to those obtained by PMFA in 3D than to 
those obtained in 2D, as expected. 
\begingroup 
\squeezetable 
\begin{table}[b] 
\caption{\label{table1} 
List of critical points $t_c/U_0$. 
}
\begin{ruledtabular}
\begin{tabular}{rccccccc}
\multicolumn{2}{c}{} &
\multicolumn{3}{c}{\textrm{Two dimensions}} &
\multicolumn{3}{c}{\textrm{Three dimensions}}\\ \cline{3-5} \cline{6-8}
$n_0$ &
$U_2/U_0$ &
$(t_c/U_0)_{\rm 3rd}$\footnote{Data obtained by third-order strong-coupling expansion} &
$t_c/U_0$\footnote{Data obtained by least-squares fit based on strong-coupling expansion. } &
$t_c/U_0$\footnote{Data obtained by chemical-potential fit.} &
$(t_c/U_0)_{\rm 3rd}{}^{\rm a}$ &
$t_c/U_0{}^{\rm b}$&
$t_c/U_0{}^{\rm c}$\\ 
\colrule
1  & 0.2 & 0.0568 & 0.0400 $\pm$ 0.0010 & 0.0482 & 0.0353 & 0.0224 $\pm$ 0.0020 & 0.0269 \\
  & 0.3 & 0.0429 & 0.0349 $\pm$ 0.0021 & 0.0365 & 0.0263 & 0.0195 $\pm$ 0.0001 & 0.0202 \\
  & 0.4 & 0.0234 & 0.0205 $\pm$ 0.0015 & 0.0201 & 0.0143 & 0.0115 $\pm$ 0.0004 & 0.0110 \\
2  & 0.2 & 0.1221 & 0.1019 $\pm$ 0.0218 & 0.1084 & 0.0758 & 0.0615 $\pm$ 0.0078 & 0.0607 \\
  & 0.3 & 0.1388 & 0.1205 $\pm$ 0.0068 & 0.1208 & 0.0866 & 0.0722 $\pm$ 0.0021 & 0.0683 \\
  & 0.4 & 0.1567 & 0.1389 $\pm$ 0.0010 & 0.1356 & 0.0978 & 0.0827 $\pm$ 0.0009 & 0.0766 \\
3  & 0.2 & 0.0313 & 0.0216 $\pm$ 0.0014 & 0.0266 & 0.0195 & 0.0121 $\pm$ 0.0016 & 0.0149 \\
  & 0.3 & 0.0236 & 0.0190 $\pm$ 0.0009 & 0.0201 & 0.0145 & 0.0106 $\pm$ 0.0001 & 0.0111 \\
  & 0.4 & 0.0129 & 0.0113 $\pm$ 0.0008 & 0.0111 & 0.0079 & 0.0063 $\pm$ 0.0002 & 0.0061 \\
4  & 0.2 & 0.0754 & 0.0611 $\pm$ 0.0150 & 0.0669 & 0.0469 & 0.0368 $\pm$ 0.0057 & 0.0375\\
  & 0.3 & 0.0859 & 0.0726 $\pm$ 0.0057 & 0.0747 & 0.0536 & 0.0435 $\pm$ 0.0021 & 0.0422\\
  & 0.4 & 0.0971 & 0.0840 $\pm$ 0.0007 & 0.0839 & 0.0606 & 0.0498 $\pm$ 0.0003 & 0.0474\\
\end{tabular}
\end{ruledtabular}
\end{table}
\endgroup

\subsection{One dimension}
In 1D, the MI phase exhibits a very rich spin structure; however, 
the strong-coupling expansion is based on the spin-singlet 
(spin-nematic) states for even (odd) boson filling. 
In particular, for odd boson filling, 
the ground state may be the spin-dimerized state over a wide parameter 
space. Thus, the results obtained by 
the proposed strong-coupling expansion cannot be directly applied to 1D 
especially for odd boson fillings. 
The range of $U_2/U_0$, in which 
we can obtain $t_{\rm C}$ at the peak of the Mott lobe, is 
more limited compared to that in 2D or 3D. For instance, 
$U_2/U_0\ge 0.255$ is required to obtain $t_{\rm C}$ for the $n_0=2$ Mott lobe.   

\begin{figure}
  \includegraphics[height=2.5in]{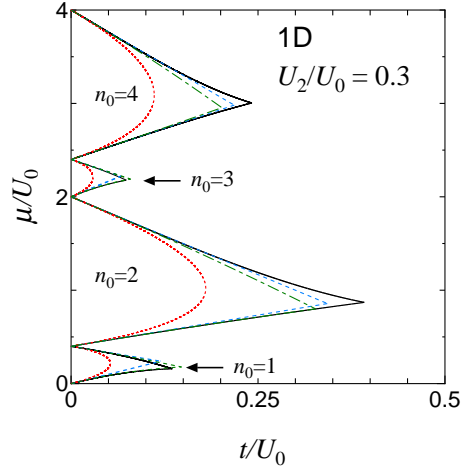}
  \caption{(Color online)
1D phase diagram at $U_2/U_0=0.3$. 
The solid curves show the results obtained by 
strong-coupling expansion up to the third order of $t$.
Results up to the first order (second order) of $t$ 
are also shown by the blue dashed (green dot-dashed) curve. 
The red dotted curve shows the results obtained by PMFA. 
}
\label{phase1Dus0.3}
\end{figure}

\begin{figure}
  \includegraphics[height=2.5in]{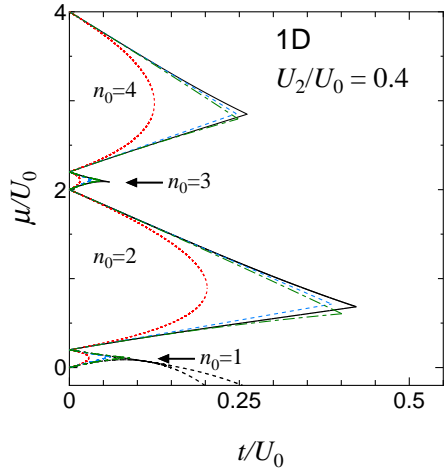}
  \caption{(Color online)
Similar plot to Fig. \ref{phase1Dus0.3} for $U_2/U_0=0.4$ in 1D, 
except for the dashed curves for the $n_0=1$ Mott lobe. 
For $n_0=1$, the upper and lower branches 
of the Mott lobe do not converge at third order in $t$, 
so a closed phase-boundary curve is not obtained. 
}
\label{phase1Dus0.4}
\end{figure}

Nevertheless, we briefly examine the phase diagram because, to date,  
most numerical simulations are for 1D. 
Figures \ref{phase1Dus0.3} and \ref{phase1Dus0.4} 
show the phase diagrams obtained by the proposed strong-coupling 
expansion up to the third order of $t$ 
for $U_2/U_0=0.3$ and $U_2/U_0=0.4$, respectively. 
In Fig. \ref{phase1Dus0.4}, the upper and lower branches  
of the $n_0=1$ Mott lobe do not converge, which 
precludes a closed phase-boundary curve. 
Figures \ref{phase1Dus0.3}  and \ref{phase1Dus0.4} also 
show that the strong-coupling expansion converges when it goes 
from the first to third order of $t$, although 
the convergence is not excellent compared with 
that in 2D or 3D 
(Figs. \ref{phasediagram2Dus015}--\ref{phasediagram3Dus03}). 

For $U_2/U_0=0.3$, agreement with the phase diagram obtained with DMRG 
\cite{Rizzi} is not excellent but satisfactory 
for the Mott lobe with even boson fillings. 
For example, with the strong-coupling expansion,  
$t_{\rm C}/U_0=0.327$, and with DMRG,  
$t_{\rm C}/U_0\simeq 0.47$ (as per our interpretation of 
Fig. 1 in Ref. \cite{Rizzi}). 
However, for $U_2/U_0=0.4$, the results do not  agree 
with the QMC results shown in Fig. 1 of Ref. \cite{Apaja}:  
 $t_{\rm C}/U_0=0.422$ using the proposed strong-coupling 
expansion and $t_{\rm C}/U_0\simeq 0.7$ from the QMC results. 
In general,  larger $t$ is required to obtain the SF phase 
for large $U_0$ and/or $U_2$, where the higher-order terms 
of $t$ become more prominent. 
Therefore, we may have to expand up to fourth or 
even higher order in order to reduce the discrepancy. 
Otherwise, we may have to assume another MI phase  
such as a nematic MI phase for even boson filling.  

Figures  \ref{phase1Dus0.3} and \ref{phase1Dus0.4} 
also show the results obtained by PMFA, 
which may be inaccurate in 1D. 
We find a very large difference between these results and 
those of the strong-coupling expansion. 

As mentioned in Sec. III F, 
we also attempt to extrapolate the results 
to infinite order in $t$. In the least-squares fit in $t_{\rm C}$,
we cannot improve the results as extensively; 
for instance, $t_{\rm C}/U_0=0.433\pm 0.030$ for $U_2/U_0=0.4$
for the $n_0=2$ Mott lobe. 
We attempt the fit with the chemical potential 
by assuming the Kosterlitz--Thouless form, as per 
Ref. \cite{Freericks}, although the fit is not successful
(figure not shown). 


\subsection{Summary and Discussion}
In this study, we used a strong-coupling expansion 
of the hopping parameter $t$ to obtain analytical 
results for the phase diagram.  
In the limit of infinite dimensions, 
the $t$--$\mu$ phase-boundary curves 
were consistent with the exact results obtained by PMFA.  

Overall, the convergence of the phase-boundary curves 
from the first to the third order was excellent.  
The dependence of $t_{\rm C}$ on $U_2$ 
at the peak of the Mott lobe with even (odd) boson filling 
indicated a reliable strong-coupling expansion at large 
(intermediate) values of $U_2$. 

We attempted to extrapolate the results to the infinite order 
in $t$ by a least-squares fit and a 
chemical-potential fit to $t_{\rm C}$. 
The linear fitting error of $t_{\rm C}$ was very small 
for large $U_2$. 
As expected, 
the fitting results of the phase-boundary curves 
agreed better with those of PMFA in 3D than with those in 2D. 
We also compared the 1D phase-boundary curves 
obtained by the strong-coupling expansion 
with those obtained by numerical simulations. 
For $U_2/U_0=0.3$, satisfactory (but not excellent) 
agreement was achieved between the 1D phase-boundary curves and 
those obtained by DMRG. 

The proposed strong-coupling expansion depends on the 
$t=0$ ground state. However, 
the MI phase can be more complicated.
We must consider the dimerized-spin phase 
for odd boson filling, which can be the ground state 
in 1D. In addition, we should consider the nematic spin phase 
for even boson filling, which can be the ground state
for weak $U_2$. To analytically determine the complete phase diagram, 
these spin phases must be included in the strong-coupling expansion. 
The possible first-order transition should also be studied
on an equal footing with the second-order transition. 
These remain problems for future work.

\appendix
\section{Energies of Mott insulator and defect states 
determined by strong-coupling expansion} 
By using $\Psi_{\rm even}$ 
and $\Psi_{\rm odd}$ [Eqs. (\ref{Me}) and (\ref{Mo})], the
energies of the MI state per site are 
\begin{eqnarray}
\frac{E_{\rm MI,even}(n_0)}{N}&=&\frac{U_0}{2}n_0(n_0-1)-U_2n_0-n_0\mu
-\frac{zt^2}{3}\frac{n_0(n_0+3)}{U_0+2U_2}, 
\label{evenenergy}\\
\frac{E_{\rm MI,odd}(n_0)}{N}&=&
\frac{U_0}{2}n_0(n_0-1)-U_2(n_0-1)-n_0\mu\nonumber\\
&&-zt^2\Big[\frac{34}{225}\frac{(n_0+4)(n_0-1)}{U_0+4U_2}
+\frac{4}{45}\frac{2n_0^2+6n_0+7}{U_0+U_2}
+\frac{1}{9}\frac{(n_0+1)(n_0+2)}{U_0-2U_2}\Big]\ \ 
\label{oddenergy}
\end{eqnarray}
up to third order in $t$. 
On the other hand, by using $\Psi^{\rm part}_{\rm even}$,
$\Psi^{\rm hole}_{\rm even}$, 
$\Psi^{\rm part}_{\rm odd}$, and 
$\Psi^{\rm hole}_{\rm odd}$ [Eqs. (\ref{pe})--(\ref{ho})], 
the energies of the defect states 
are 
\begin{eqnarray}
&&E^{\rm part}_{\rm def,even}(n_0)-E_{\rm MI,even}(n_0)
\nonumber\\
&=& U_0n-\mu-zt\frac{n_0+3}{3}\nonumber\\
&& -\frac{z(z-7)t^2}{9}\frac{n_0(n_0+3)}{U_0+2U_2}
-\frac{zt^2n_0}{9}\Big[2\Big(\frac{n_0+5}{2U_0+3U_2}
            +\frac{n_0+3}{3U_2}\Big)
            +\frac{n_0+2}{2U_0}\Big]\nonumber\\
&&-\frac{zt^3}{27}n_0(n_0+3)
        \Big\{
        (z-1)\Big[
           \frac{(2n_0+3)z-3(3n_0+8)}{(U_0+2U_2)^2}\nonumber\\
&&           +\frac{2}{U_0+2U_2}\Big(
           2\frac{n_0+5}{2U_0+3U_2}+\frac{n_0+2}{2U_0}\Big)
           +\frac{4(n_0+3)}{3U_2(U_0+2U_2)}\Big]\nonumber\\
&&        -z\Big[
          2\Big(\frac{n_0+5}{(2U_0+3U_2)^2}+\frac{n_0+3}{(3U_2)^2}\Big)
          +\frac{n_0+2}{(2U_0)^2}\Big]
        +\frac{4}{3U_2}\Big(
        \frac{1}{5}\frac{n_0+5}{2U_0+3U_2}+\frac{n_0+2}{2U_0}\Big)
        \Big\},\label{evenpart}
\end{eqnarray}
\begin{eqnarray}
&&E^{\rm hole}_{\rm def,even}(n_0)-E_{\rm MI,even}(n_0)\nonumber\\
&=&-U_0(n_0-1)+2U_2+\mu-zt\frac{n_0}{3}\nonumber\\
&&-\frac{z(z-7)t^2}{9}\frac{n_0(n_0+3)}{U_0+2U_2}
-\frac{zt^2(n_0+3)}{9}\Big[2\Big(\frac{n_0-2}{2U_0+3U_2}
            +\frac{n_0}{3U_2}\Big)
            +\frac{n_0+1}{2U_0}\Big]\nonumber\\
&&-\frac{zt^3}{27}n_0(n_0+3)
        \Big\{
        (z-1)\Big[
           \frac{(2n_0+3)z-3(3n_0+1)}{(U_0+2U_2)^2}\nonumber\\
&&           +\frac{2}{U_0+2U_2}\Big(
           2\frac{n_0-2}{2U_0+3U_2}+\frac{n_0+1}{2U_0}\Big)
           +\frac{4n_0}{3U_2(U_0+2U_2)}\Big]\nonumber\\
&&        -z\Big[
          2\Big(\frac{n_0-2}{(2U_0+3U_2)^2}+\frac{n_0}{(3U_2)^2}\Big)
          +\frac{n_0+1}{(2U_0)^2}\Big]
        +\frac{4}{3U_2}\Big(
        \frac{1}{5}\frac{n_0-2}{2U_0+3U_2}+\frac{n_0+1}{2U_0}\Big)
        \Big\},\label{evenhole}
\end{eqnarray}
\begin{eqnarray}
&&E^{\rm part}_{\rm def,odd}(n_0)-E_{\rm MI,odd}(n_0)\nonumber\\
&=&U_0n_0-2U_2-zt\frac{n_0+1}{3}-\mu\nonumber\\
&&-\frac{z(z-3)t^2}{9}(n_0+1)\Big[\frac{n_0+2}{U_0-2U_2}
          +\frac{4}{5}\frac{n_0-1}{U_0+U_2}\Big]\nonumber\\
&&-\frac{zt^2}{9}(n_0+4)\Big[2\frac{n_0-1}{2U_0+3U_2}
                    +\frac{n_0+2}{2U_0}
-\frac{68}{25}\frac{n_0-1}{U_0+4U_2}
-\frac{8}{5}\frac{n_0+2}{U_0+U_2}\Big]\nonumber\\
&&-\frac{2}{45}z(2z+3)t^2\frac{(n_0+1)(n_0+4)}{3U_2}\nonumber\\
&&-\frac{z(z-1)t^3}{27}\frac{(n_0+1)(n_0+2)}{(U_0-2U_2)^2}
                     \big[(2n_0+3)z-(5n_0+6)\big]\nonumber\\
&&-\frac{4}{675}z(z-1)t^3\frac{n_0+1}{(U_0+U_2)^2}
                     \big[(n_0-1)(9n_0+1)z-2(17n_0^2+26n_0+32)\big]\nonumber\\
&&-\frac{z(z-1)^2t^3}{27}\frac{n_0+1}{U_0+U_2}
\Big[\frac{32}{25}\frac{(n_0-1)(n_0+4)}{U_0+4U_2}
     +\frac{8}{5}\frac{(n_0+2)(2n_0+3)}{U_0-2U_2}\Big]\nonumber\\
&&-\frac{2}{27}z(z-1)t^3(n_0+1)(n_0+4)
\Big\{\frac{n_0-1}{2U_0+3U_2}\Big[\frac{34}{25}\frac{1}{U_0+4U_2}
             +\frac{4}{5}\frac{1}{U_0+U_2}\Big]\nonumber\\
&&      +\frac{n_0+2}{2U_0}\Big[\frac{1}{U_0-2U_2}
             +\frac{4}{5}\frac{1}{U_0+U_2}\Big]
      +\frac{1}{3U_2}\Big[\frac{2}{25}(8z+9)\frac{n_0-1}{U_0+4U_2}
             +\frac{4}{5}z\frac{n_0+2}{U_0+U_2}\Big]\Big\}\nonumber\\
&&-\frac{4}{135}z(2z+3)t^3\frac{(n_0+1)(n_0+4)}{3U_2}
      \Big[\frac{1}{5}\frac{n_0-1}{2U_0+3U_2}+\frac{n_0+2}{2U_0}\Big]\nonumber\\
&&+\frac{2}{675}zt^3\frac{(n_0+1)(n_0+4)}{(3U_2)^2}
      \big[2(n_0-11)z^2+9(3n_0+7)z-9(n_0+4)\big]\nonumber\\
&&+\frac{zt^3}{27}(n_0+1)(n_0+4)\Big[
        \frac{68}{25}(z-1)\frac{n_0-1}{(U_0+4U_2)^2}
        +2z\frac{n_0-1}{(2U_0+3U_2)^2}+z\frac{n_0+2}{(2U_0)^2}\Big],
\label{oddpart}
\end{eqnarray}
\begin{eqnarray}
&&E^{\rm hole}_{\rm def,odd}(n_0)-E_{\rm MI,odd}(n_0)\nonumber\\
&=&-U_0(n_0-1)-zt\frac{n_0+2}{3}+\mu\nonumber\\
&&-\frac{z(z-3)t^2}{9}(n_0+2)\Big[\frac{n_0+1}{U_0-2U_2}
          +\frac{4}{5}\frac{n_0+4}{U_0+U_2}\Big]\nonumber\\
&&-\frac{zt^2}{9}(n_0-1)\Big[2\frac{n_0+4}{2U_0+3U_2}
                    +\frac{n_0+1}{2U_0}
-\frac{68}{25}\frac{n_0+4}{U_0+4U_2}
-\frac{8}{5}\frac{n_0+1}{U_0+U_2}\Big]\nonumber\\
&&-\frac{2}{45}z(2z+3)t^2\frac{(n_0-1)(n_0+2)}{3U_2}\nonumber\\
&&-\frac{z(z-1)t^3}{27}\frac{(n_0+1)(n_0+2)}{(U_0-2U_2)^2}
                     \big[(2n_0+3)z-(5n_0+9)\big]\nonumber\\
&&-\frac{4}{675}z(z-1)t^3\frac{n_0+2}{(U_0+U_2)^2}
                     \big[(n_0+4)(9n_0+26)z-2(17n_0^2+76n_0+107)\big]\nonumber\\
&&-\frac{z(z-1)^2t^3}{27}\frac{n_0+2}{U_0+U_2}
\Big[\frac{32}{25}\frac{(n_0-1)(n_0+4)}{U_0+4U_2}
     +\frac{8}{5}\frac{(n_0+1)(2n_0+3)}{U_0-2U_2}\Big]\nonumber\\
&&-\frac{2}{27}z(z-1)t^3(n_0-1)(n_0+2)
\Big\{\frac{n_0+4}{2U_0+3U_2}\Big[\frac{34}{25}\frac{1}{U_0+4U_2}
             +\frac{4}{5}\frac{1}{U_0+U_2}\Big]\nonumber\\
&&      +\frac{n_0+1}{2U_0}\Big[\frac{1}{U_0-2U_2}
             +\frac{4}{5}\frac{1}{U_0+U_2}\Big]
      +\frac{1}{3U_2}\Big[\frac{2}{25}(8z+9)\frac{n_0+4}{U_0+4U_2}
             +\frac{4}{5}z\frac{n_0+1}{U_0+U_2}\Big]\Big\}\nonumber\\
&&-\frac{4}{135}z(2z+3)t^3\frac{(n_0-1)(n_0+2)}{3U_2}
      \Big[\frac{1}{5}\frac{n_0+4}{2U_0+3U_2}+\frac{n_0+1}{2U_0}\Big]\nonumber\\
&&+\frac{2}{675}zt^3\frac{(n_0-1)(n_0+2)}{(3U_2)^2}
      \big[2(n_0+14)z^2+9(3n_0+2)z-9(n_0-1)\big]\nonumber\\
&&+\frac{zt^3}{27}(n_0-1)(n_0+2)\Big[
        \frac{68}{25}(z-1)\frac{n_0+4}{(U_0+4U_2)^2}
        +2z\frac{n_0+4}{(2U_0+3U_2)^2}+z\frac{n_0+1}{(2U_0)^2}\Big]
\label{oddhole}
\end{eqnarray}
up to third order in $t$. 
By equating the right-hand side of Eqs. (\ref{evenpart})--(\ref{oddhole})
with zero, we obtain the SF--MI phase-boundary curve.

\section{Expansion of phase-boundary curve obtained by PMFA}
The phase-boundary curves obtained by PMFA are 
given by Eqs. (30) and (46) of Ref. \cite{Tsuchiya}
for even and odd MI lobes, respectively. 
We can straightforwardly expand the results 
up to third order in $zt$ for even MI lobes:
\begin{eqnarray}
\mu^{\rm part}_{\rm even}&=&U_0n_0-\frac{n_0+3}{3}zt
  -\frac{n_0(n_0+3)}{9}\frac{(zt)^2}{U_0+2U_2}\nonumber\\
  &&-\frac{n_0(n_0+3)(2n_0+3)}{27}\frac{(zt)^3}{(U_0+2U_2)^2}\label{muevenpart}\\
\mu^{\rm hole}_{\rm even}&=&U_0(n_0-1)-2U_2+\frac{n_0}{3}zt
  +\frac{n_0(n_0+3)}{9}\frac{(zt)^2}{U_0+2U_2}\nonumber\\
  &&+\frac{n_0(n_0+3)(2n_0+3)}{27}\frac{(zt)^3}{(U_0+2U_2)^2}
\label{muevenhole}
\end{eqnarray}
and for odd MI lobes,  
\begin{eqnarray}
\mu^{\rm part}_{\rm odd}&=&U_0n_0-2U_2-\frac{n_0+3}{3}zt
  -\frac{n_0+1}{9}\Big[\frac{n_0+2}{U_0-2U_2}
    +\frac{4}{5}\Big(\frac{n_0-1}{U_0+U_2}+\frac{n_0+4}{3U_2}\Big)\Big](zt)^2
 \nonumber\\
&&  -\frac{n_0+1}{27}\Big\{
       \frac{(n_0+2)(2n_0+3)}{(U_0-2U_2)^2}
       +\frac{4}{25}\frac{(n_0-1)(9n_0+1)}{(U_0+U_2)^2}
       -\frac{4}{25}\frac{(n_0+4)(n_0-11)}{(3U_2)^2}
 \nonumber\\
&&       +\frac{24}{25}\frac{(n_0+4)(3n_0+2)}{3U_2(U_0+U_2)}
       +\frac{8}{5}\frac{(n_0+2)(2n_0+3)}{(U_0+U_2)(U_0-2U_2)}\Big\}(zt)^3
\label{muoddpart}\\
\mu^{\rm hole}_{\rm odd}&=&U_0(n_0-1)+\frac{n_0+2}{3}zt
  +\frac{n_0+2}{9}\Big[\frac{n_0+1}{U_0-2U_2}
    +\frac{4}{5}\Big(\frac{n_0-1}{3U_2}+\frac{n_0+4}{U_0+U_2}\Big)\Big](zt)^2
 \nonumber\\
&&  +\frac{n_0+2}{27}\Big\{
       \frac{(n_0+1)(2n_0+3)}{(U_0-2U_2)^2}
       +\frac{4}{25}\frac{(n_0+4)(9n_0+26)}{(U_0+U_2)^2}
       -\frac{4}{25}\frac{(n_0-1)(n_0+14)}{(3U_2)^2}
 \nonumber\\
&&           +\frac{24}{25}\frac{(n_0-1)(3n_0+7)}{3U_2(U_0+U_2)}
             +\frac{8}{5}\frac{(n_0+1)(2n_0+3)}{(U_0+U_2)(U_0-2U_2)}\Big\}(zt)^3.
\label{muoddhole}
\end{eqnarray}


\begin{thebibliography}{99}
\bibitem{comment0}
For instance, by a similar calculation to that of Ref. \cite{Kimura1},  
we obtain that the first-order SF--MI transition 
from the Mott lobe with two bosons per site 
disappears when $U_2/U_0>0.32$ in the Gutzwiller MF study.
\bibitem{comment2}
The critical exponent is determined by 
phase fluctuation on the fixed-density contour,  
so we consider that the only relevant degree of freedom 
is the SF phase, as for the spinless BH model. 
In addition, at $U_2=0$, 
the critical exponent must be the same as that of the spinless 
BH model because the spin magnetic quantum number 
of all bosons must be the same ($S_z=0,\pm 1$ are degenerate) in order 
to maximize the absolute value of  kinetic energy, 
and the spin degree of freedom does not influence the critical behavior. 
Even if $U_2$ becomes finite (but sufficiently small) 
and the degeneracy is lifted, the polar state 
consisting of only $S_z=0$ bosons 
(without $S_z=\pm 1$) remains the ground state. 
The critical behavior usually does not depend on the 
value of microscopic parameters, so 
the critical behavior for large $U_2$ is probably 
the same as that for $U_2\simeq 0$.
\end{thebibliography}
\end{document}